\documentclass{mn2e}
\usepackage{epsfig}
\usepackage{times}

\newcommand{\hh}{^{\rm h}}
\newcommand{\mm}{^{\rm m}}

\title{X-ray emission from the nuclei, lobes and hot-gas environments of two FR-II radio galaxies} 
\author[J. H. Croston et al.]
       {J. H. Croston, \thanks{Email: Judith.Croston@bris.ac.uk} 
	M. Birkinshaw, 
	M. J. Hardcastle, 
	and D. M. Worrall \\
        H. H. Wills Physics Laboratory, University of Bristol, Tyndall Avenue, Bristol BS8 1TL }

\pagerange{\pageref{firstpage}--\pageref{lastpage}}
\pubyear{2004}
\begin{document}

\maketitle

\label{firstpage}

\begin{abstract}

We report the detection of multiple components of X-ray emission from
the two FR-II radio galaxies 3C~223 and 3C~284, based on new {\it
XMM-Newton} observations. We attribute the detected X-ray emission
from the lobes of both sources to inverse-Compton scattering of cosmic
microwave background photons. With this model, we find that the magnetic
field strength in the lobes is at the equipartition value for 3C~284,
and within a factor of two of the equipartition value for 3C~223. We
also detect group-scale hot atmospheres around both sources, and
determine temperatures and pressures in the gas. The lobes of both
sources are in pressure balance with the hot-gas environments, if the
lobes contain only the synchrotron-emitting particles and the measured
magnetic field strength. The core spectra of both sources contain an
unabsorbed soft component, likely to be related to the radio jet, and
an additional heavily absorbed power-law component. 3C~223 also
displays a bright (EW $\sim$ 500 eV) Fe K$\alpha$ emission line.

\end{abstract}


\section{Introduction}

X-ray observations are providing answers to many long-standing
questions about the dynamics and physical conditions of radio
galaxies. Studies of radio-galaxy environments have revealed the
influence of the hot-gas medium on radio-lobe structure and have shown
that radio-source heating may be common (Kraft et al. 2003; Croston et
al. 2003). Detections of X-ray synchrotron and inverse-Compton (IC)
jets have constrained the electron population in the jets (e.g. Harris
\& Krawczynski 2002) and provided information about the locations of
particle acceleration in low-power (FR-I) jets (e.g. Hardcastle et al.
2003). Measurements of IC emission from lobes and hotspots are
providing information about the magnetic field strength and the
relative contribution of particles and field to source dynamics (e.g.
Brunetti et al. 2001, Hardcastle et al. 2002a; Isobe et al. 2002).

Prior to the availability of sensitive X-ray data, studies of
radio-galaxy dynamics and source evolution were based almost entirely
on measurements of radio synchrotron emission. As synchrotron
emissivity depends on magnetic field strength as well as electron
energy density, our knowledge of the physical properties of the
sources has been limited. The assumption of equipartition of energy
density between particles (relativistic electrons) and magnetic field
has typically been used to obtain values for the physical properties
of jets, lobes and hotspots. However, there are problems with this
assumption. In particular, the internal radio-lobe pressures
determined on this basis have been found to be significantly lower
than the external pressures inferred from X-ray measurements in both
FR-I and FR-II radio galaxies (e.g. Morganti et al. 1988; Worrall \&
Birkinshaw 2000; Hardcastle \& Worrall 2000a), which is inconsistent
with lobe expansion. Investigation of the validity of this assumption
is therefore essential for a full understanding of the dynamics and
evolution of radio galaxies.

X-ray IC emission has been detected from the lobes and hotspots of a
number of radio galaxies (e.g. Harris et al. 1994; Hardcastle et al.
2001; Brunetti et al. 2001; Hardcastle et al. 2002a; Belsole et al.
2004). As this emission process does not involve the magnetic field,
the X-ray IC flux can be used in combination with the radio
synchrotron spectrum to determine the magnetic field strength of the
emitting region. To date, detected lobe IC sources are typically found
to be near to equipartition or in the particle-dominated regime, with
magnetic fields $B_{ic} = (0.1 - 1) \times B_{eq}$, where $B_{eq}$ is
the equipartition value. However, the results are dependent on the
parameters of the input electron energy spectrum used to model the
radio synchrotron emission, and on the careful separation of thermal
and non-thermal X-ray emission, so that estimates from different
authors are often not directly comparable, and the overall picture
remains unclear.

The study of lobe IC emission also has implications for the dynamics
of radio galaxies. As defined above, the equipartition calculation
assumes that the only particles in the radio lobes are the
relativistic electrons (and positrons) responsible for the radio
emission. If the lobes contain large contributions from additional
particles, such as relativistic protons, then there is no physical
reason for them to prefer magnetic field strengths close to the value
of $B_{eq}$ calculated for electrons only. Therefore, if observations
show that sources {\it are} commonly found with $B \sim B_{eq}$, this
suggests that the minimum-energy lobe pressures are correct and that
there is no energetically dominant population of non-radiating
particles. By combining observations of lobe-related emission and
thermal emission from the surrounding medium, it is possible to test
models of radio-source expansion.

It is therefore useful to find more examples of radio galaxies with
lobe-related X-ray emission. In this paper, we describe new {\it
XMM-Newton} observations of two powerful (FR-II) radio galaxies, 3C
223 ($z = 0.1368$) and 3C 284 ($z = 0.2394$), and discuss the dynamics
of the radio sources and their impact on their surrounding hot-gas
environments.

We use a cosmology with H$_{0}$ = 70 km s$^{-1}$ Mpc$^{-1}$,
$\Omega_{M}$ = 0.3, and $\Omega_{\Lambda}$ = 0.7 throughout, which
gives a scale of 2.4 kpc arcsec$^{-1}$ at the distance of 3C 223, and
a scale of 3.8 kpc arcsec$^{-1}$ at the distance of 3C 284.

\section{Data reduction and analysis}

{\it XMM-Newton} observed 3C~223 in October 2001 and 3C~284 in
December 2002. Both observations were made using the medium filter,
and the pn data were taken in extended full frame mode. For 3C~223,
the duration was 33903 s for the MOS 1 camera, 33917 s for the MOS 2
camera and 25849 s for the pn camera, and for 3C~284, the duration was
43066 s for MOS1, 43104 s for MOS2 and 35806 s for pn.

We reduced the data using the {\it XMM-Newton} Scientific Analysis
Software (SAS) package, using methods described in the XMM-SAS
Handbook. The data were filtered for good time intervals using a
count-level threshold determined by examining a histogram of the count
rate above 10 keV. For 3C 223, the threshold levels used were 4 cts/s
for MOS1, 3 cts/s for MOS2, and 5 cts/s for pn; for 3C 284 they were
1.4 cts/s for MOS1 and MOS2 and 1.3 cts/s for pn.  The data were then
filtered using the flag bitmask 0x766a0600 for MOS and 0xfa000c for
pn, which are equivalent to the standard flagset \#XMMEA\_EM/EP but
include out of field-of-view events (useful for studying the particle
component of the background) and exclude bad columns and rows. They
were also filtered for patterns less than or equal to 12 for the MOS
cameras and less than or equal to 4 for the pn, as suggested in the
Handbook. The 3C~223 observation included several large background
flares, so that the filtered data were of duration 23138s for MOS1,
24651 s for MOS2 and 12305 s for pn. The filtered 3C~284 data were of
duration 41775 s for MOS1, 41496 s for MOS2 and 32095 s for pn.

Our images were created using software that interpolates over the {\it
XMM-Newton} chip gaps (described in more detail in Croston et al.
2003) to avoid artefacts in smoothed images. We then adaptively
smoothed the images using the {\sc ciao} task {\it csmooth} for the
purpose of better identifying contaminating point sources. The point
sources were removed from the unsmoothed images using the {\sc
ciao} task {\it dmfilth}, and the region files retained for use in
spectral analysis. None of the point sources coincided with the
positions of the radio lobes, so that we cannot have accidentally
removed any radio-related emission. The resulting images were smoothed
using Gaussian kernels in order to show the distribution of extended
emission, and also adaptively smoothed to show compact structure in
the regions of extended emission. As the X-ray structure shown in the
images is on-axis, we do not include the vignetting correction in the
images. See Section 2.1 for a more detailed discussion of the vignetting
correction.

Spectral analysis was performed using scripts based on the {\sc sas}
{\it evselect} tool to extract spectra from all three cameras. For the
spectral analysis, we included the vignetting correction by using {\sc
sas} task {\it evigweight}, so as to be able to use on-axis response
files downloaded from the {\it XMM-Newton} website rather than
indivually generated responses for each spectrum and camera. We checked
that there is no significant difference in the results we obtain with
these response files and with files generated using {\it rmfgen}. We
generated ancillary response files using {\it arfgen}. As the
extraction regions are typically small and on-axis, we used local
background subtraction, because this is likely to be a more accurate
representation of the true background in the source regions than
background spectra obtained from template files. The source and
background spectra were scaled to account for differences in area
using the {\sc sas} task {\it backscale}, as {\it evselect} does not
put this information in spectrum headers. Fig.~\ref{223reg} shows the
choice of spectral regions for 3C~223 and Fig.~\ref{284reg} for
3C~284. We used {\sc xspec} for spectral model fitting, and all of our
fits assume Galactic absorption, with $N_{H} = 1.4 \times 10^{20}$
cm$^{-2}$ and $N_{H} = 9.9 \times 10^{19}$ cm$^{-2}$ (Dickey \&
Lockman 1990) for 3C~223 and 3C~284, respectively.

In order to search for extended emission that might be associated with
a hot-gas atmosphere, we used radial surface-brightness profiling of
the central regions. Software developed in earlier work (e.g.
Birkinshaw \& Worrall 1993; Hardcastle et al. 2002b) was used to
extract and fit radial profiles.  We fitted models obtained by
convolving a point source with a $\beta$-model to the
surface-brightness profiles, and compared these with point-source only
models to study the significance of any contribution from extended
emission. For the point-source models we used the analytical
description in the document XMM-SOC-CAL-TN-0022, obtained from the
{\it XMM-Newton} website.

\subsection{Issues with the vignetting correction and particle background}

The inclusion of the vignetting correction has the effect of
incorrectly weighting up unvignetted particle events. By examining the
events in the region outside the field of view, we found that the
contribution of particle events to the background is particularly high
in these two datasets (between 30 and 50 per cent of events in the
background spectra), so that these incorrectly weighted events could
significantly affect our results.

For imaging the X-ray structure in our dataset, it is preferable not
to include the correction, because the weighted-up particle events
lead to bright regions at the edges of the image, particularly after
smoothing. For spectral analysis, the correction is more important.
However, our extraction regions are mainly on-axis, so that the
weighting factor is typically quite small (always less than 20 per
cent). We concluded that the effect of including the correction, which
leads to a slight overestimation of the background level for spectral
analysis, is less than the error introduced by not correcting the
spectra, which would lead to an underestimation of the background
level. Therefore, all of our spectral analysis was performed with the
vignetting-corrected events list.

This compromise means that systematic errors are introduced into the
background spectra at a level of $\sim 7$ per cent for 3C~223 and
$\sim 11$ per cent for 3C~284. We compared corrected and uncorrected
spectra for our largest extraction region for each source, and found
consistent best-fitting spectral parameters, but a significant
difference in the model flux.  Where it could affect our conclusions,
we calculated a systematic error in flux measurements by determining
the fractional increase in background-subtracted source counts if the
background is over-estimated by the amounts given above. These provide
conservative upper limits on the flux values, as source counts are
also slightly overestimated (always by less than 2 per cent, since the
weighting factor is lower in the source regions).

We included the vignetting correction in the radial-profile analysis,
for similar reasons as above. Since the low surface brightness of the
extended emission leads to large uncertainties on $\beta$-model
parameters, the error introduced here by the weighted-up particle
events is not significant.

\section{Results}

\begin{table*}
\caption{0.3 -- 7.0 keV background-subtracted counts extracted for each spectral region.}
\label{counts}
\centering
\begin{tabular}{l|llll}
\hline
&\multicolumn{2}{c}{3C~223}&\multicolumn{2}{c}{3C~284}\\
Region&Counts&Area (arcmin$^{2}$)&Counts&Area (arcmin$^{2}$)\\
\hline
core&$1836\pm43$&3.1&$929\pm31$&0.35\\
N/E lobe&$231\pm15$&2.3&$123\pm11$&0.73\\
S/W lobe&$148\pm12$&2.0&$82\pm8$&0.75\\
extended&304$\pm$17&17.9&332$\pm$18&13.3\\
\hline
\end{tabular}
\end{table*}

\subsection{3C 223}

\begin{figure}
\epsfig{figure=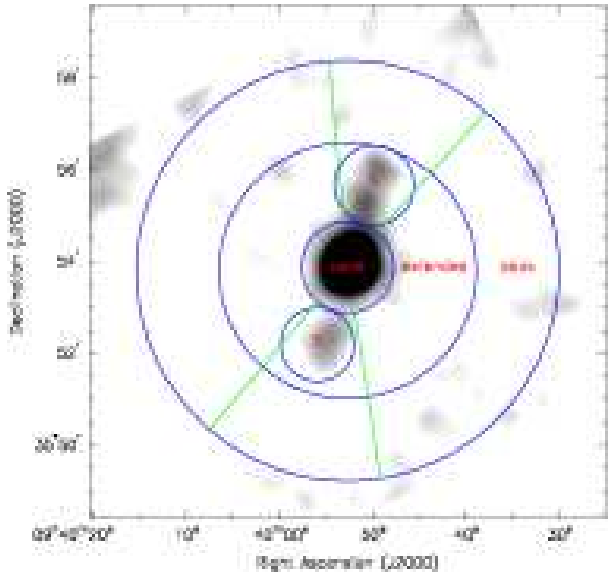,width=8.0cm}
\caption{Extraction regions used to study the core (to a radius of
  1$^{\prime}$ so as to include the wings of the XMM PSF), lobes and
  extended emission of 3C~223. Note that the extended region excludes
  the core and position angles between 148 and 190$^{\circ}$, and
  between 320 and 370$^{\circ}$ to avoid contamination from
  lobe-related emission. We used local background (the ``extended''
  region) for the lobe spectra. The extended emission, which is a
  small fraction of the background level, has been smoothed out in
  this image to show the structure of the core and lobes.}
\label{223reg}
\end{figure}

\begin{figure*}
\centerline{\hbox{
\epsfig{figure=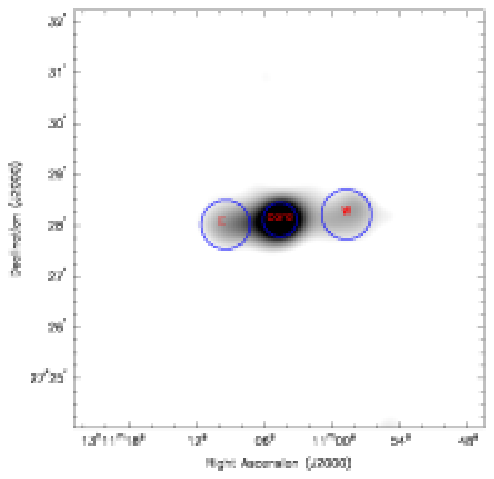,width=7.0cm}
\epsfig{figure=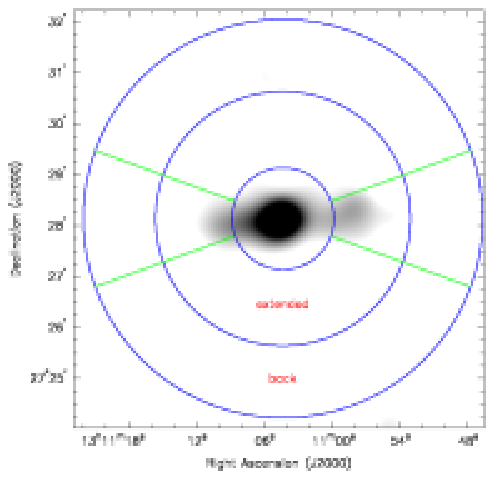,width=7.0cm}}}
\caption{Extraction regions used to study the core and lobes (left)
  and extended emission (right) of 3C~284. To obtain background
  spectra for the lobes, the two extraction regions were rotated by 90
  degrees. As with 3C~223, the extended emission is only a small
  fraction of the background level, so that it has been smoothed away
  in this image to show the core and lobe structure.}
\label{284reg}
\end{figure*}

\begin{figure}
\epsfig{figure=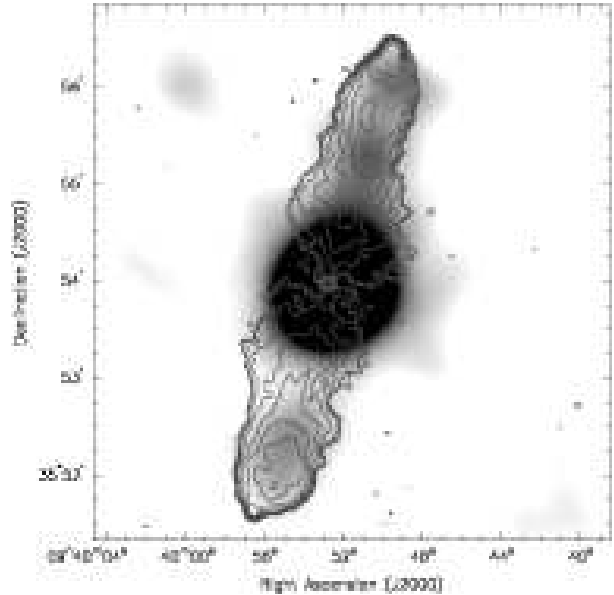,width=8.0cm}
\caption{A Gaussian smoothed ($\sigma$ = 3.5 arcsec), background point-source subtracted
  image in the energy range 0.3 -- 7.0 keV made from the combined
  MOS1, MOS2 and pn data for 3C~223, with 1.4-GHz radio contours (from
  the 4 arcsec resolution VLA map of Leahy \& Perley 1991) overlaid, showing X-ray components associated with the north and south lobes, the core, and an extended environment. Contour levels are ($\sqrt{2}$,2,4,...,128) $\times$ 1.8 $\times 10^{-5}$ Jy/beam.}
\label{223im}
\end{figure}

\begin{figure}
\epsfig{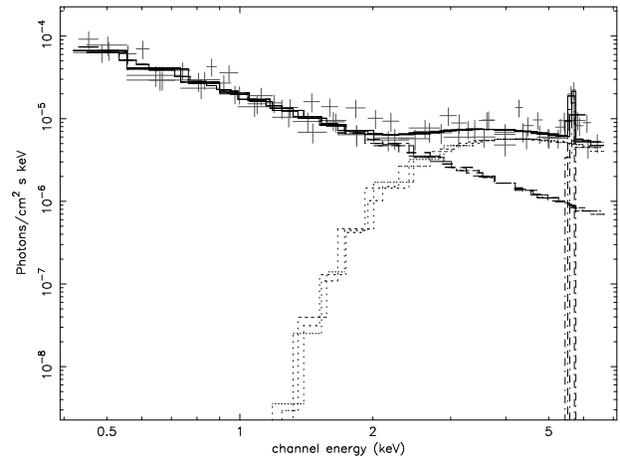}
\caption{Combined MOS1, MOS2 and pn core spectrum for 3C~223, with
  Model I, described in the text.}
\label{223core}
\end{figure}

\begin{figure*}
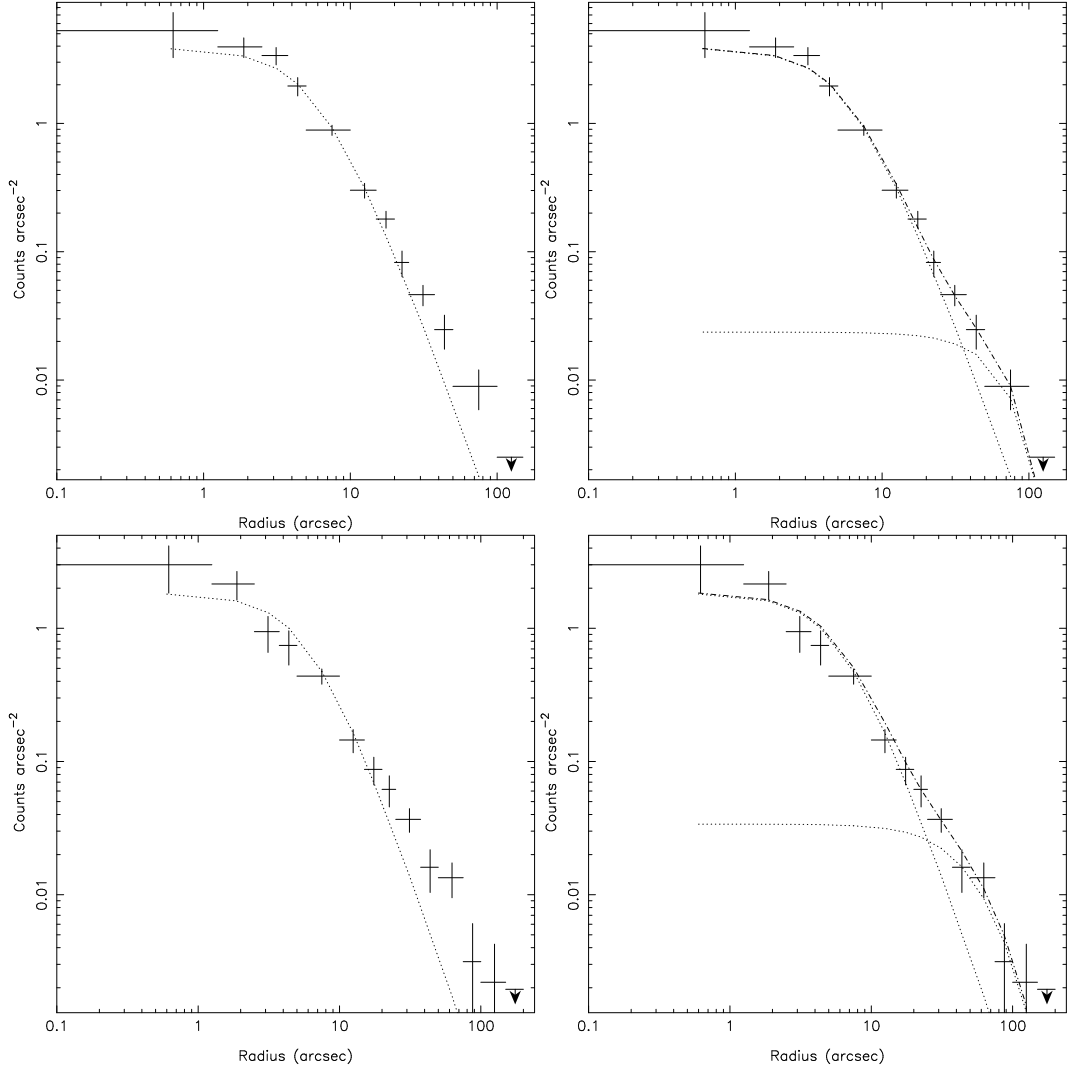

\centerline{\vbox{\hbox{
\epsfig{figure=223_pn1mod.ps,width=7.0cm}
\epsfig{figure=223_pn2mod.ps,width=7.0cm}}
\hbox{
\epsfig{figure=284_pn2beta.ps,width=7.0cm}
\epsfig{figure=284_pn1beta.ps,width=7.0cm}}}}
\caption{Radial surface brightness profiles for 3C~223 and 3C~284 (pn
  data). The top plots are the single point-source model (left) and
  point-source plus $\beta$ model (right) for 3C~223; the bottom plots
  are the the same for 3C~284. The $\beta$-model component has been
  convolved with the PSF determined as described in the text.}
\label{radprof}
\end{figure*}

Fig.~\ref{223im} shows a smoothed image of 3C~223 with radio contours
overlaid. The most prominent features in the X-ray emission are the
core and lobe-related emission. However, the extended region around
the core suggests the presence of a hot-gas atmosphere.

We extracted spectra from the MOS1, MOS2 and pn files, as described
above, to study the core and lobe emission. The choice of extraction
regions is illustrated in Fig.~\ref{223reg}, and the counts in each
region are given in Table~\ref{counts}. We initially fitted the core
spectrum with a single power law model, but this is an unacceptable
fit ($\chi^{2}$ of 179 for 70 d.o.f.), so that multiple components are
needed. We fitted two multiple-component models to the continuum:
Model I, consisting of a soft, unabsorbed power law, and a hard,
absorbed power law; and Model II, consisting of a soft, thermal
component with fixed abundance of 0.3 solar, and a hard, absorbed
power law. In addition, to model residuals at around 6 keV, we
included a redshifted Gaussian component of fixed linewidth 10 eV. We
assumed Galactic absorption as given in Section 2. It was necessary to
fix the power-law index of the hard component, so as to constrain the
fit. We chose a value of 1.5, which is the expected value for IC
nuclear scattering; however, the choice of $\Gamma$ does not
significantly affect the fit. The best-fitting model parameters are
shown in Table~\ref{components}. Both models are good fits to the
data, although Model I gives a slightly better fit statistic. The
inclusion of the Gaussian component results in a significant
improvement in the fit statistic. We performed an F-test for
comparison with other results in the literature, which shows that the
improvement is significant at $>$ 99.99 per cent confidence level;
however, we note that, although commonly used, F-tests are not
reliable for this purpose (Protassov et al. 2002). We allowed the
linewidth of the Gaussian to vary and found a best-fitting value of
$176^{+54}_{-45}$ eV, which is slightly larger than, but consistent
with the expected pn spectral resolution at this energy, so that there
is no strong evidence for broadening of the line. We measured a
rest-frame equivalent width for the 6.4 keV line of 517$^{+99}_{-97}$
eV. Fig.~\ref{223core} shows Model I fitted to the MOS1, MOS2 and pn
spectra. We discuss the physical origin of the model components in
Section 4.1.

\begin{table*}
\caption{Models for 3C~223's core spectrum. Fluxes are in units of
  ergs s$^{-1}$ cm$^{-2}$.}
\label{components}
\begin{tabular}{ll|rrrr}
\hline 
Model component&Parameter&Model I&Model II\\ 
\hline 
Soft, unabsorbed power law&$\Gamma$&$2.01\pm0.11$&\\ 
&Unabsorbed flux (0.3 -- 7 keV)&$(1.3\pm0.1) \times 10^{-13}$&\\ 
Soft {\it mekal}&$kT$(keV)&&$1.62^{+0.12}_{-0.07}$\\ 
&Unabsorbed flux (0.3 -- 7 keV)&&$(8.6^{+0.4}_{-0.5}) \times
10^{-14}$\\ 
Nuclear column density&$N_{H}$ (cm$^{-2}$)&$(9.5\pm1.5) \times 10^{22}$&$(7.5\pm1) \times
10^{22}$\\ 
Hard, absorbed power law&$\Gamma$&1.5 (frozen)&1.5 (frozen)\\
&Unabsorbed flux (0.3 -- 7 keV)&$(5.5\pm0.5) \times 10^{-13}$&$(6.0\pm0.5) \times 10^{-13}$\\ 
Gaussian ($z=0.1368$)&Energy(keV)&$6.43^{+0.07}_{-0.05}$&$6.43^{+0.07}_{-0.05}$\\ 
&Line width (eV)&10 (frozen)&10 (frozen)\\ 
&Unabsorbed flux (0.3 -- 7 keV)&$(2.5\pm0.5) \times 10^{-14}$&$(2.7^{+0.3}_{-0.7} \times 10^{-14}$\\
$\chi^{2}/dof$&&85/80&81/80\\ 
\hline
\end{tabular}
\end{table*}

We then studied the emission associated with the radio lobes of
3C~223, regions N and S in Fig.~\ref{223reg}. We initially fitted the
lobe spectra with {\it mekal} and power-law models. The parameters for
these fits are shown in Table~\ref{lobespec}. The models are equally
acceptable fits to the data, but the best-fitting temperatures for the
thermal models are high. As we have used local background subtraction,
the spectra should not be contaminated by emission from the hot-gas
environment. We adopt the power-law models, which seem more physically
plausible than the hot {\it mekal} models; however, we discuss an
interpretation based on the {\it mekal} fits in Sections 4.2 and 4.5.
For the power-law model, the unabsorbed 1-keV flux densities are
$3.1\pm0.6$ nJy and $3.0\pm0.5$ nJy for the north and south lobes,
respectively. We discuss models for the lobe emission in further
detail in Section 4.2.

We used radial surface-brightness profiling to search for extended
emission around the radio sources. As there are insufficient counts in
the two MOS cameras to fit a profile successfully, we used only the pn
events file for this analysis. We extracted the counts in annuli
centred on the source, excluding angles where lobe emission is present
as marked in Fig.~\ref{223reg}. Fig~\ref{radprof} shows the profile
with a point-source model (left) and with this plus an additional
$\beta$-model component (right). The inclusion of the $\beta$-model
component results in an improved fit. To determine the statistical
significance of the improvement, we carried out Monte Carlo
simulations using the PSF model for the pn camera and counting
statistics. We then fitted the two models to the fake datasets and
determined the F-statistic for each set of fits; these were compared
with the measured value of F (28.5). We find a less than 1 per cent
probability that the improvement in the fit could occur by chance. The
best-fitting $\beta$-model parameters are $\beta = 1.5$ and $r_{c}$ =
140 arcsec, but the parameters are very poorly constrained, so that a
very good fit can be obtained for any plausible lower value of $\beta$.

To confirm the presence of extended emission and investigate its
nature, we extracted spectra from an annulus of inner radius 60 arcsec
and outer radius 167.5 arcsec using the radial profiles to determine
the location of the expected extended emission, and excluding the
angles where the radio lobe emission is present and several point
sources not associated with the radio source. The radial profiles for
3C~223 show that we expect a significant fraction of the flux in our
extraction region to be scattered emission from the point source. From
the encircled energy fraction information in the PSF document
described in Section 2, we would expect 7 per cent of the core
spectrum to be scattered into the extraction annulus. The energy
dependence of the PSF is not important at these radii and so the
scattered spectrum will not be significantly altered. We therefore
included fixed components in the model for the extended spectrum,
corresponding to our adopted core model described above, and
normalised to 7 per cent of the core counts. We then fitted the
temperature and normalisation of a {\it mekal} component with 0.3
solar abundance, holding the core components fixed. Table~\ref{223ex}
shows the best-fitting {\it mekal} parameters for the {\it mekal} plus
scattered point-source model. The unabsorbed fluxes of the extended
and scattered point-source components (not given in the Table) are
roughly consistent with the ratios shown in the radial profile. We
estimated the systematic error caused by incorrectly weighted particle
events, as described in Section 2.1. In this case it is very large
($\sim 300$ per cent), due to the small number of source counts
relative to background in the region; however, even a factor of 3
increase in flux does not affect our later analysis and conclusions,
as shown in Section 4.4.

\subsubsection{Nearby quasar}

\begin{figure}
\begin{center}
\epsfig{figure=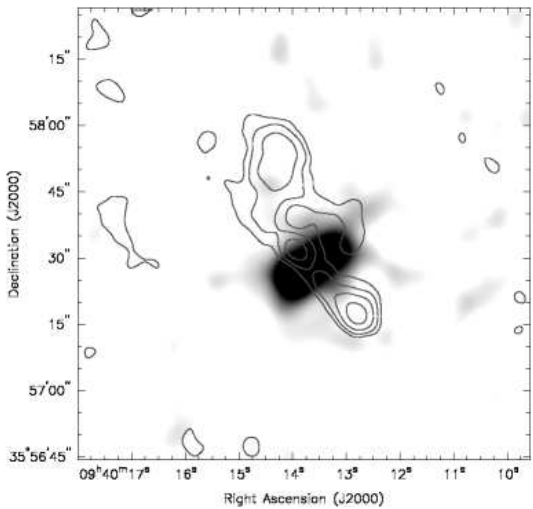,width=7.0cm}
\caption{A Gaussian smoothed ($\sigma = 2$ arcsec) combined MOS1, MOS2
  and pn image of the X-ray-detected quasar described in Section
  3.1.1. Radio contours overlaid are from the same map used in
  Fig.~\ref{223im}. There is slight misalignment of the data so that
  the X-ray and radio cores are offset by $\sim$ 1 arcsec.}
\label{baby}
\end{center}
\end{figure}

Our radio map for 3C~223 shows a small double-lobed radio source 5.6
arcmin to the north-east of the source, at (J2000)
RA~$09\hh40\mm13\fs9$, Dec~$+35\degr57'32''$. We find a bright X-ray
source at that position in our {\it XMM-Newton} data. Fig.~\ref{baby}
shows the radio contours overlaid on a Gaussian smoothed image of the
combined MOS1, MOS2 and pn X-ray data. We extracted a radial profile
of the X-ray data, which shows that it is consistent with being a
point source. We extracted an X-ray spectrum in a circle of radius 47
arcsec, and find a good fit to the spectrum ($\chi^{2}=18$ for 19
d.o.f.) with a power-law model, having photon index $\Gamma =
1.92\pm0.18$. We measure a 0.3 - 7.0 flux of $(5.4\pm0.5) \times
10^{-14}$ ergs cm$^{-2}$ s$^{-1}$ (1-keV flux density of 7.2$\pm$0.7
nJy).

This X-ray source was previously detected with {\it ROSAT} and an
optical identification with a star was made (Mason et al. 2000).
However, the star is offset by 20 arcsec from the X-ray and radio
source. We obtained the DSS2 images for this field, and find that in
additional to the star of the previous optical identification, there
is a faint object at exactly the position of the X-ray source in the
blue image, but none in the red. We therefore conclude that the X-ray
source was previously misidentified, and is in fact a distant quasar.

\begin{table*}
\centering
\caption{Best-fitting parameters of the spectral models for the lobe
  regions of 3C~223 and 3C~284. Errors are 1$\sigma$ for one
  interesting parameter. The abundance was fixed at 0.3 solar for all
  {\it mekal} fits.}
\label{lobespec}
\begin{tabular}{ll|rrrr}
\hline
&&\multicolumn{2}{c}{3C 223}&\multicolumn{2}{c}{3C 284}\\
&&North&South&East&West\\
\hline
Power law (I)&$\Gamma$&1.4$\pm$0.3&1.0$\pm$0.5&2.1$\pm$0.3&1.2$\pm$0.6\\
&1-keV flux density (nJy)&3.1$\pm$0.5&3.0$\pm$0.5&1.9$\pm$0.2&0.9$^{+0.1}_{-0.2}$\\
&$\chi^{2}$ (dof)&11 (9)&6 (5)&3 (8)& 2 (2)\\
{\it mekal} (II)&$kT$ (keV)& $>$5&$>$6&2.3$^{+1.0}_{-0.5}$&$>$3\\
&0.3 -- 7.0 keV flux (ergs cm$^{-2}$ s$^{-1})$&3.$\times10^{-14}$&3.7$\times10^{-14}$&1$\times10^{-14}$&1$\times10^{-14}$\\
&$\chi^{2}$ (dof)&11 (9)& 6 (5)& 5 (8)& 2 (2)\\
\hline
\end{tabular}
\end{table*}

\begin{table}
\caption{Parameters of the {\it mekal} component for the fit to the
  extended emission surrounding 3C~223. Fluxes are measured in units
  of ergs cm$^{-2}$ s$^{-1}$, and luminosities in ergs s$^{-1}$. The
  flux and luminosity in the model come from the spectral regions
  described in the text, and so exclude the core and angles where
  there is lobe emission. The bolometric total luminosity is
  calculated by integrating the $\beta$-model fits to surface
  brightness over the entire model to a radius of 150 arcsec, and so
  includes the core and lobe regions which were excluded from the
  spectral fit for the extended region.}
\label{223ex}
\centering
\begin{tabular}{l|rr}
\hline
$kT$ (keV)& 1.4$^{+2.9}_{-0.5}$\\
Flux in model (0.3 -- 7.0 keV)& ($2.7\pm1.1) \times 10^{-14}$\\
Luminosity in model (0.3 -- 7.0 keV)&$(1.3\pm0.5) \times 10^{42}$ \\
$\chi^{2}$ (dof)& 4.5 (7) \\
Bolometric total luminosity&$8.5^{+246}_{-3.6} \times 10^{42}$\\
\hline
\end{tabular}
\end{table}

\subsection{3C~284}

\begin{figure}
\epsfig{figure=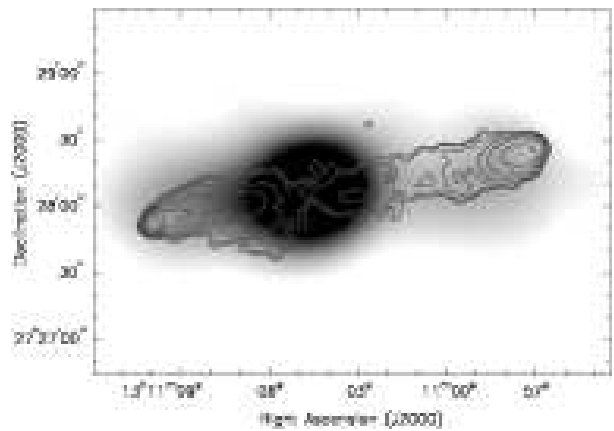,width=8.0cm}
\caption{A Gaussian smoothed ($\sigma$ = 14 arcsec), background point-source subtracted
  image made from the combined MOS1, MOS2 and pn data in the 0.3 --
  7.0 keV band for 3C~284, with 1.4-GHz radio contours (made using
  data taken from the VLA archive) overlaid. The radio map has
  5-arcsec resolution and contour levels are ($\sqrt{2}$,2,4,...,128)
  $\times$ 1.2 $\times 10^{-3}$ Jy beam$^{-1}$.}
\label{284im}
\end{figure}

\begin{figure}
\epsfig{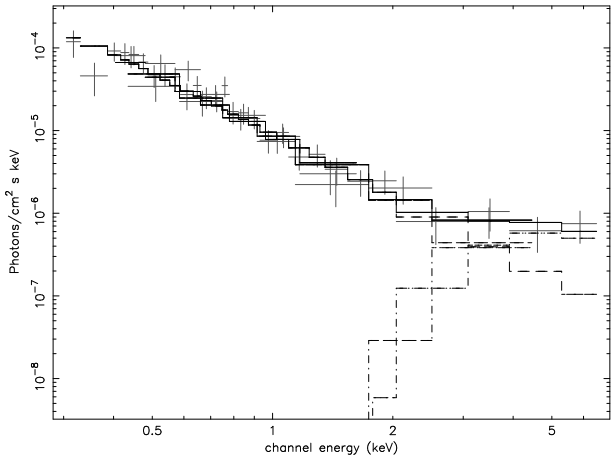}
\caption{Core spectrum for 3C~284, with the best-fitting 2-component power-law
  model, as described in the text.}
\label{284core}
\end{figure}

In Fig.~\ref{284im}, we show an adaptively smoothed image of the X-ray
emission associated with this source, with radio contours overlaid. As
with 3C~223, the most prominent X-ray features appear to be associated
with the core and radio lobes. We studied core and lobe spectra and
used radial surface-brightness profiling to search for extended
emission, as above. Fig.~\ref{284reg} shows our choice of extraction
regions. For this source, we used a smaller core extraction circle, so
as to avoid contamination from the lobe emission. We fitted a single
power-law model, as well as two multiple-component models as for
3C~223. For the soft {\it mekal} plus hard, absorbed power-law model
(similar to Model II for 3C~223), we found that the column density
tended to zero, so that we instead fitted a {\it mekal} plus
unabsorbed power law. Table~\ref{284_core} gives the best-fitting
parameters for the three models. Model I is a single power-law fit,
Model II a soft power law plus hard, absorbed power law (similar to
Model I for 3C~223), and Model III is a {\it mekal} plus power-law
model. All three models are accpetable fits to the data, although II
and III give lower values of $\chi^{2}$. The best-fitting temperature
for Model III is much lower than that of the extended atmosphere (see
later), and the luminosity is much higher than would be expected if it
was simply the inner regions of the extended environment. We therefore
adopt Model II. The spectrum with this best-fitting two power-law
model is shown in Figure~\ref{284core}. There are insufficient counts
at high energies to determine whether there are any emission lines.

\begin{table*}
\caption{Best-fitting parameters for the three models fitted to the
  3C~284 core spectrum}
\label{284_core}
\centering
\begin{tabular}{llrrr}
\hline
Model component&Parameter&Model I&Model II&Model III\\
\hline
Soft, unabsorbed power law&$\Gamma$&$2.44^{+0.10}_{-0.11}$&$2.54\pm0.11$&$1.91^{+0.18}_{-0.15}$\\
&Unabsorbed flux (0.3 - 7.0 keV)&$(4.4^{+0.3}_{-0.2}) \times 10^{-24}$&$(4.3^{+0.3}_{-0.2})
\times 10^{-14}$&$(3.8^{+0.5}_{-0.4}) \times 10^{-14}$\\
Soft {\it mekal}&$kT$ (keV)&&&$0.32\pm0.07$\\
&Unabsorbed flux (0.3 - 7.0 keV)&&&$(9\pm3) \times 10^{-15}$\\
Nuclear column density&N$_{H}$ (cm$^{-2}$)&&($26^{+14}_{-10}) \times 10^{22}$&\\
Hard, absorbed power law&$\Gamma$&&1.5 (frozen)&\\
&Unabsorbed flux (0.3 - 7.0 keV)&&$(7.1^{+8.4}_{-3.7})
\times 10^{-14}$&\\
$\chi^{2}/dof$&&41/40&33/38&32/38\\
\hline
\end{tabular}
\end{table*}

As was done for 3C~223, we fitted the lobe-region spectra with single
{\it mekal} and power-law models. For the eastern lobe, the power-law
model is a slightly better fit (see Table~\ref{lobespec}); however,
for the western lobe, both models fit adequately. We discuss
interpretations based on each model in Section 4. For the power-law
model, we measured 1-keV flux densities of 1.9$\pm$0.2 nJy and
0.9$^{+0.1}_{-0.2}$ nJy for the east and west lobes, respectively.
Table~\ref{lobespec} shows the best-fitting parameters for the lobe
emission models. 

Radial profiles for 3C~284 are shown in Fig.~\ref{radprof}. The
profiles were extracted from the pn events file, excluding the
position angles where there is lobe emission, and excluding point
sources. We show both point-source and point-source plus $\beta$
models. The inclusion of the $\beta$-model component produces a
significant improvement in the fit. The best-fitting $\beta$-model
parameters are $\beta = 0.75$ and $r_{c}$ = 56 arcsec; however, the
parameters are poorly constrained, so we do not quote errors. We used
the same method as for 3C~223 to examine the significance of the
inclusion of a $\beta$-model component, and find there is a less than
1 per cent probability of such an improvement (F = 45.7) occurring by
chance.

To study the extended emission in the regions identified by the radial
profiles, we extracted spectra from an annulus of inner radius 60
arcsec and outer radius 150 arcsec excluding the angles where the
radio lobe emission is present and any contaminating background point
sources. We fitted the spectra with a single {\it mekal} model, as the
radial profile (Fig~\ref{radprof}) shows that the emission in our
extraction region is dominated by the atmosphere. The best-fitting
{\it mekal} parameters are shown in Table~\ref{284ex}. As with 3C~223,
the systematic error from incorrectly weighted particle events is
high, $\sim 300$ per cent, since the ratio of source to background
counts is low. This does not have a significant effect on our later
conclusions, as shown in Section 4.4.

\begin{table}
\caption{Parameters of the best-fitting {\it mekal}-model fits to the
  extended emission surrounding 3C~284. Fluxes are measured in units
  of ergs cm$^{-2}$ s$^{-1}$, and luminosities in ergs s$^{-1}$. The
  flux and luminosity in the model come from the spectral regions
  described in the text, and so exclude the core and angles where
  there is lobe emission. The bolometric total luminosity is
  calculated by integrating the $\beta$-model fits to surface
  brightness over the entire model out to a radius of 150 arcsec, and
  so includes the core and lobe regions which were excluded from the
  spectral fit for the extended region.}
\label{284ex}
\centering
\begin{tabular}{l|rr}
\hline
$kT$ (keV)& 1.03$^{+0.36}_{-0.20}$ \\
Flux in model (0.3 -- 7.0 keV)& $(1.4\pm0.4) \times 10^{-14}$\\
Luminosity in model (0.3 -- 7.0 keV)& $(2.4\pm0.7) \times 10^{42}$\\
$\chi^{2}$/n (dof)&11 (13)\\
Bolometric total luminosity&$(3.8^{+5.5}_{-1.0}) \times 10^{42}$\\
\hline
\end{tabular}
\end{table}

\section{Discussion}

\subsection{Nuclear spectra}

The two models we fitted to the nuclear spectrum of 3C~223 lead to two
physical explanations for the soft emission from the core. If we adopt
Model I, the soft power-law component of the nuclear spectrum could be
associated with the base of the radio jet, as it must originate
outside the dusty material obscuring the second component. Hardcastle
\& Worrall (1999) report a correlation between radio and X-ray core
flux based on {\it ROSAT} measurements, which suggests that a large
fraction of the soft X-ray core flux may be associated with the
small-scale radio jets. Our measured flux from the primary power law
is in agreement with the Hardcastle \& Worrall relation. If we adopt
the slightly better fit of Model II, then the thermal material has a
temperature consistent with the extended thermal atmosphere; however,
its luminosity is too high for this to be its origin. For 3C~284, a
two-power law model, similar to Model I for 3C~223, is preferred.
Therefore we can explain the nuclear spectra of both sources with a
radio-related soft power-law and an absorbed, hard nuclear power law.
Our results are consistent with those found for the nuclear spectrum
of Cygnus A (Young et al. 2002), although they find a slightly better
fit with a thermal origin for the soft emission, and also consistent
with other radio-galaxy nuclear spectra (e.g. Hardcastle et al. 2002a;
Belsole et al. 2004).

As discussed in Section 3.1, for the core spectrum of 3C~223, we find
strong evidence for a redshifted 6.4-keV line in all three cameras.
Although we find a best-fitting line width of $176^{+54}_{-45}$ eV, the
spectral resolution of the XMM cameras (e.g. 130 eV FWHM at 6.4 keV
for the pn camera) is comparable to this value, so that there is no
evidence for line broadening. Our measured equivalent width of
517$^{+99}_{-96}$ eV is within the range of measured values for
powerful radio galaxies ($\sim$ 100 -- 1500 eV; see e.g. Sambruna et
al. 1999). However, only 4 objects in their sample out of 39 (3
broad-line objects, and one narrow-line radio galaxy) have equivalent
widths greater than 450 eV. 3C~223 is a narrow-line radio galaxy, and
the majority of iron-line detections in FR-IIs have been in broad-line
radio galaxies, where the source geometry is expected to be similar to
that of quasars. However, the brightest iron line so far detected (in
3C~321: EW 1510 eV, Sambruna et al. 1999) is in a narrow-line object.
Ptak et al. (1996) present theoretical predictions for the relation
between equivalent width and column density of material in which the
fluorescence is occuring, for several models of source geometry. The
column density of absorbing material we find to be obscuring the
second power-law component in our core spectra is $\sim$ 10$^{23}$
cm$^{-2}$, so that our results are roughly consistent with the
relations described in Ptak et al. (1996). The most plausible
explanation for the bright iron line in 3C~223 is that we are seeing
fluorescence of cold gas in a thick torus obscuring the nucleus of
3C~223.

\subsection{Lobe emission from the two radio galaxies}

In neither source was the spectral fitting able to distinguish between
the power-law and {\it mekal} models for the lobe-related emission.
For 3C~223, the best-fitting {\it mekal} temperatures are high,
and in both cases they are poorly constrained. One possible origin for
such hot gas would be in a model where the supersonic lobe expansion
is shock-heating the environment. We discuss the arguments for and
against this model in Section 4.5, but in the remainder of this
Section we adopt the single power-law model for the lobes of both sources,
assuming the X-rays to be radio-related.

\begin{figure*}
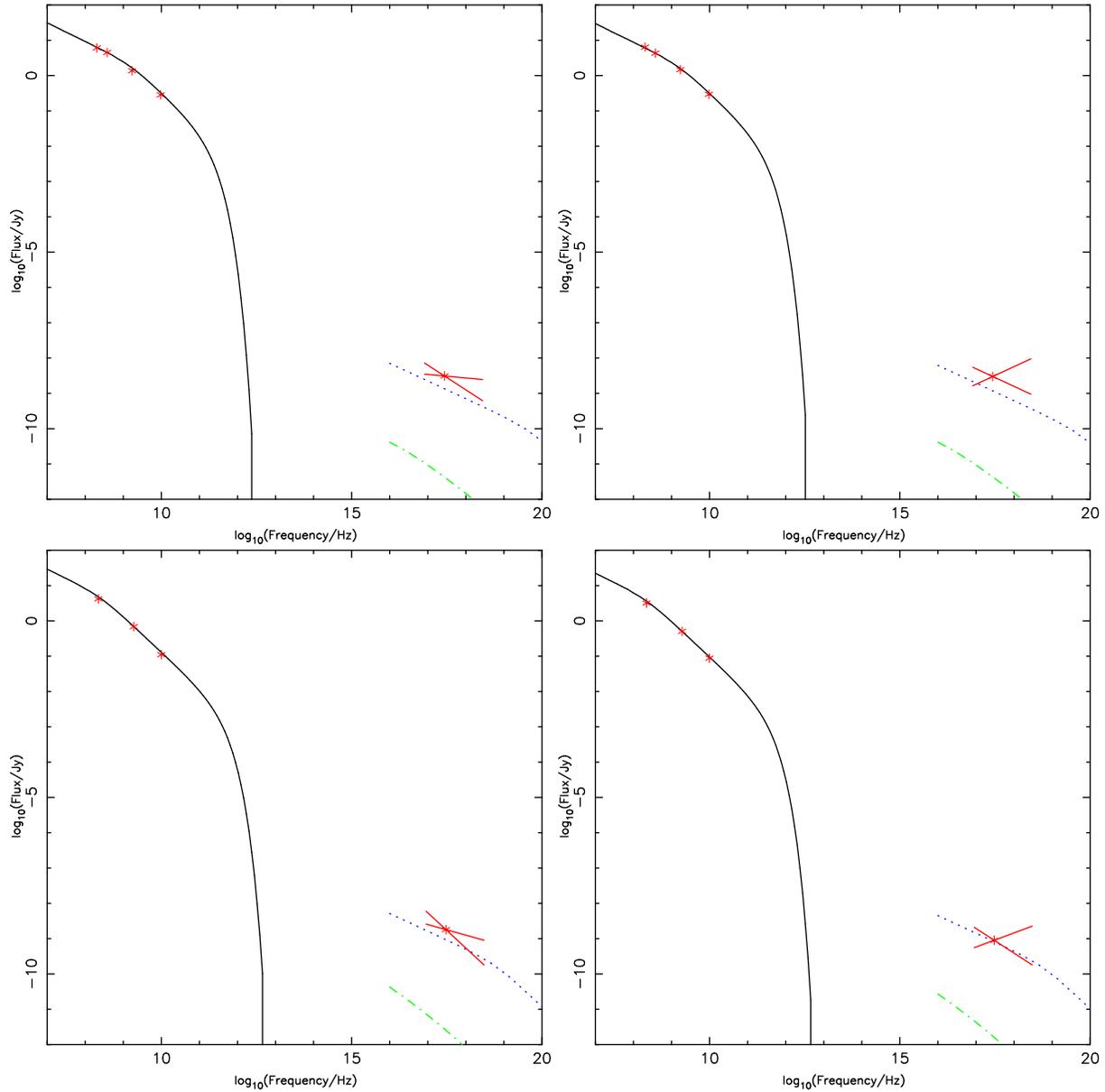

\centerline{\vbox{\hbox{
\epsfig{figure=223_nlobe.ps,width=8.0cm}
\epsfig{figure=223_slobe.ps,width=8.0cm}}
\hbox{
\epsfig{figure=284_elobe.ps,width=8.0cm}
\epsfig{figure=284_wlobe.ps,width=8.0cm}}}}
\caption{Radio-to-X-ray spectra for the north and south lobes of
3C~223 (top) and the east and west lobes of 3C~284 (bottom), with
synchrotron models fitted to the radio data points as described in the
text, and predicted CMB inverse-Compton (dotted line) and SSC (dashed
line) at the position of the X-ray flux, showing that in all cases the
measured flux is close to the predicted level for CMB inverse-Compton
emission at equipartition, and the spectral indices are also
consistent with this interpretation.}
\label{lobes}
\end{figure*}

In Fig.~\ref{lobes}, we plot the radio-to-X-ray spectra for the two
radio lobes of each source with an equipartition inverse-Compton model
determined using the code of Hardcastle et al. (1998). Radio fluxes
were obtained at 1.5 and 8 GHz from maps of Leahy \& Perley (1991),
and Hardcastle et al (1998). The average of the fraction of the total
flux in each lobe at the two measured frequencies was used to
determine 178-MHz fluxes from the 3CRR fluxes (Laing, Riley \& Longair
1983). For 3C~223, we also included 330 MHz fluxes from data obtained
from the VLA archive and reduced in the standard manner. A synchrotron
spectrum was fitted to the radio data with an inital energy index of
2, minimum energy of $5 \times 10^{6}$ eV, maximum of $6 \times
10^{11}$ eV, and break of 1 in energy index at an energy of $6 \times
10^{9}$ eV for 3C~223 and $2 \times 10^{9}$ eV for 3C~284, assuming a
filling factor of unity. The spectral break parameters were chosen so
as to obtain a good fit to the radio data. The synchrotron models and
predicted inverse-Compton fluxes for each lobe from cosmic microwave
background (CMB) photons and from synchrotron self-Compton (SSC)
emission, are shown in Fig.~\ref{lobes}.

The equipartition magnetic field strengths (assuming no relativistic
protons) were determined from the synchrotron model above to be $\sim$
0.35 nT for both the north and south lobes of 3C~223, and 0.5 nT for
the east and west lobes of 3C~284, respectively. Fig.~\ref{lobes}
shows that in all cases the measured flux is fairly close to the value
predicted for inverse-Compton scattering of CMB photons by an electron
population at equipartition. If the measured X-ray flux values are
used to determine the value for $B$, the magnetic field strength, we
obtain values of 0.22 nT and 0.20 nT for the north and south lobes of
3C~223, which are within a factor of two of the equipartition values.
In 3C~284, the measured magnetic field strength is 0.4 nT for the
eastern lobe, and the same as the equipartition value for the western lobe.

\begin{table}
\caption{Predicted 1-keV flux density from nuclear inverse Compton
  emission in the eastern lobe of 3C~284. Column 1 is the lobe's angle to the
  line-of-sight. Columns 2 and 3 give the predicted X-ray flux density
  for two values of the low-energy cut-off to the electron energy spectum.}
\label{nic}
\centering
\begin{tabular}{lrr}
\hline
$\theta$ (degrees)&\multicolumn{2}{c}{1-keV flux density (nJy)}\\ 
& $\gamma_{min} = 10$&$\gamma_{min} = 100$\\
\hline
90&0.053&0.046\\
105&0.068&0.060\\
120&0.074&0.068\\
135&0.068&0.062\\
\hline
\end{tabular}
\end{table}

The X-ray emission from the eastern lobe of 3C~284 is brightest near
the core and decreases slightly with radius out to a distance of 45
arcsec (170 kpc), whereas the radio emission increases with distance.
This could be explained if there was a significant contribution to the
X-ray emission from inverse Compton scattering of a nuclear photon
field (e.g. Brunetti et al. 1997). In addition, the X-ray to radio
ratio in the eastern lobe is higher than in the western lobe (they
differ at $>$ 2$\sigma$ level), which could be explained by the
anisotropy of IC emission, which leads to more emission in the
direction towards the illuminating source, if the eastern lobe is the
more distant one. We therefore computed the expected flux from IC
scattering by the eastern radio lobe of infrared and optical photons
from a hidden quasar, using the results of Brunetti (2000) as
described by Hardcastle et al. (2002a). The electron energy population
was modelled in the same way as for the SSC and CMB calculations
above; however, the calculations were repeated for a low-energy
cutoff, $\gamma_{min}$, of 10 to include electrons of sufficiently low
energy to scatter the optical photons to the X-ray region. The
illuminating point source was modelled using the SED of the quasar
3C~273 (of similar radio luminosity and redshift to 3C~284). We used
flux density measurements at 180 $\mu$ (Meisenheimer et al. 2001), 1
$\mu$ (Spinoglio et al. 1995), and U-band (Neugebauer et al. 1979) to
parametrize the infrared to optical spectrum. We performed the
calculation for angles to the line-of-sight of 105, 120 and 135
degrees to determine the maximum possible contribution to the X-ray
flux. The results of the calculations are given in Table~\ref{nic}.
The highest flux value of 0.074 nJy is obtained for an angle of 120
degrees and $\gamma_{min} = 10$, so that for all choices of angle to
the line-of-sight the total contribution from nuclear IC to the X-ray
flux is less than 5 per cent of the observed flux. In this best-case
scenario, the ratio of observed to predicted IC flux for the eastern
and western lobes agree at the 2$\sigma$ level. We therefore conclude
that there could be some contribution from nuclear IC emission to the
eastern lobe; however, the dominant photon population comes from the
CMB. The contribution from nuclear IC emission in the case of 3C~223
is expected to be even less, as the radio luminosity is lower by a
factor of three, and the X-ray emission follows the radio structure
more closely, as expected in the CMB model.

\subsection{Group atmospheres}

For both 3C~223 and 3C~284, the combination of spectral analysis and
radial profile analysis leads to an unambiguous detection of an
extended atmosphere. 3C~284 possesses a group-scale atmosphere
(L$_{X}$ $\sim 4 \times 10^{42}$ ergs s$^{-1}$) with a temperature of
$\sim$ 1 keV. This temperature is close to that expected for
a group of this luminosity from the L$_{X}$/T relation for radio-quiet
groups (Croston et al. 2003). We also produced a hardness-ratio
map for 3C~284, but there was no evidence for hotter gas around the
lobes; due to the small number of counts from the environment, it was
not possible to separate the harder lobe emission from any hot gas
emission near the lobe edges. 

3C~223 has an atmosphere of luminosity $\sim 9 \times 10^{42}$ ergs
s$^{-1}$, giving an expected temperature of 1.2 keV. This is
consistent with the measured temperature of $1.4^{+2.9}_{-0.5}$ keV.
As with 3C~284, there are insufficient counts to establish whether or
not there are temperature variations in the atmosphere. We conclude
that there is no evidence for large-scale heating due to the radio
source in the atmosphere of either source.

\subsection{Pressure balance}

\begin{table*}
\caption{Internal and external pressure measurements for 3C~223 and
  3C~284. The $R$ values are the ratios of internal to external
  pressure}
\label{pressures}
\centering
\begin{tabular}{llrrrr}
\hline
&P$_{int}$/Pa&P$_{ext}$ (midpoint)/Pa&P$_{ext}$/Pa (end)&R$_{mid}$&R$_{end}$\\
\hline
3C~223N& 4.0$\times 10^{-14}$&$3.1^{+2.0}_{-1.3} \times 10^{-13}$&$9.6^{+38.3}_{-8.8} \times 10^{-14}$&0.13&0.42\\
3C~223S& 5.3 $\times 10^{-14}$&$3.1^{+2.0}_{-1.3} \times 10^{-13}$&$9.6^{+38.3}_{-8.8} \times 10^{-14}$&0.17&0.55\\
3C~284E& 8.3 $\times 10^{-14}$&$1.4^{+0.2}_{-0.4} \times 10^{-13}$ &$6.6^{+1.0}_{-2.1} \times 10^{-14}$&0.6&1.3\\
3C~284W& 6.0 $\times 10^{-14}$&$1.0^{+0.1}_{-0.2} \times 10^{-13}$&$3.8^{+1.7}_{-2.8} \times 10^{-14}$&0.6&1.6\\
\hline
\end{tabular}
\end{table*}

We determined internal lobe pressures from the radio-synchrotron
models described in Section 4.2, above, assuming an electron filling
factor of unity, and that the only contribution to pressure comes from
the population of relativistic electrons producing the synchrotron
emission. We used the magnetic field strengths given above calculated
based on the measured IC flux. We then calculated the external
pressure on the lobes from the X-ray-emitting gas using the the
best-fitting thermal models for the extended emission described in
Section 4.3, above. We used the results of Birkinshaw \& Worrall
(1993) to determine pressure as a function of radius from the emission
measure and the $\beta$-model fit parameters. For each source, we
determined the pressure halfway along the lobes, and at the ends of
the lobes. For 3C~284, we did the pressure calculation separately for
each lobe as they are asymmetrical in length.

Internal and external pressures for each source are given in
Table~\ref{pressures}. The errors we quote for pressures are based on
the errors on the best-fitting $\beta$-model parameters, which are the
main contribution to uncertainty in the pressure. We assume that
$\beta$ is between 0.35 and 2, as the fits are poorly constrained. For
3C~223, the factor of three uncertainty in the flux from the
atmosphere leads to a $\sim 50$ per cent increase in the pressure
values (not included here), which does not significantly affect our
conclusions below about pressure balance. For 3C~284, the factor of
three uncertainty in the flux from the atmosphere leads to a 50 -- 70
percent increase in the pressure values (not included here). As for
3C~223, this does not significantly affect the conclusions.

Table~\ref{pressures} shows that 3C~223 is consistent with being in
pressure balance at the ends of the source, but likely to be
underpressured at the midpoint. For 3C~284 there is also approximate
pressure balance at the ends of the source, and at the midpoint the
lobes are underpressured by a factor of $\sim$ 2. These results
suggest that self-similar models of radio-source expansion (e.g. Falle
1991; Kaiser \& Alexander 1997; see also a discussion of these models
in Hardcastle \& Worrall 2000a), which require that the entire source
be overpressured, no longer apply to these sources. 3C~223 and 3C~284
are comparatively large sources, so that it seems plausible that they
have reached a stage of expansion where the inner parts of the sources
are now underpressured. The outer parts of the source are still
expanding, but the lateral expansion is now subsonic. Further support
for this model comes from the radio structure of the two sources,
which have a ''pinched'' appearance in the central regions. Gas that
was intially pushed out by lobe expansion may now be falling back in
and crushing the central part of the cocoon.

If we instead adopt the second interpretation of the lobe emission --
that it is thermal in origin, and hot -- then the lateral expansion of
the lobes must be supersonic and overpressured, so as to produce
shock-heated gas surrounding the entire lobes. In this scenario, since
all of the X-ray flux from the lobes is attributed to hot gas, the
level of IC emission from the lobes must be much lower than the values
we used in Section 4.2. This means that the lobes must either be
magnetically dominated, so that the pressure for supersonic expansion
comes from the magnetic field, or else contain a dominant proton or
other non-radiating particle contribution. This interpretation
requires two coincidences: an X-ray flux which happens to be at the
level expected for the minimum energy condition in the absence of
protons, and that the minimum energy condition happens to correspond
to an internal pressure approximately in balance with the pressure
from the surrounding hot gas.

We therefore consider that the IC model for the lobe emission is far
more plausible. However, in the following section, we discuss the
constraints that our results have placed on models that require
overpressured lobes.

\subsection{Constraints on supersonically expanding lobes}

In order for supersonic lateral expansion of the lobes, they must
either be magnetically dominated, or dominated by non-radiating
particles, as described above. The lobes of low-power (FR-I) radio
galaxies do require a large contribution from one of these
constituents; however, it seems likely that much of the additional
pressure in FR-Is comes from material entrained into the lobes as the
jets decelerate (e.g. Bicknell 1984, Laing \& Bridle 2002). In FR-IIs,
where entrainment is not thought to be an efficient process, a
different explanation is probably necessary if the lobes are required
to be overpressured. It is therefore tempting to accept the
self-consistent model we proposed in the previous section, in which
the two sources are at equipartition and in approximate pressure
balance, so as to avoid requiring this additional explanation.
Nevertheless, we briefly consider possible pressure contributions to
FR-II lobes in the following paragraphs, both to constrain
supersonically expanding models, and since our upper limits on
external pressure for 3C~223 leave room for a large contribution from
other particles in that source.

Constraints on the amount of cold gas and a subrelativistic tail to
the electron population can be obtained by examining the limits on
Faraday depolarization. For 3C~284, Leahy, Pooley \& Riley (1986)
report only slight depolarisation between 11 and 18cm. To investigate
whether 3C\,223 showed evidence for internal or external
depolarization we retrieved VLA data from the public archive. The
low-frequency data were from 1.5-GHz observations in the VLA B and C
configurations, described by Leahy \& Perley (1991), while the
high-frequency data we used were from the two-pointing 8.4-GHz
observations at C and D configuration described by Leahy et al.
(1997). The total and polarized intensities were compared within a
fixed contour (the 5-$\sigma$ level of the 8.4-GHz total intensity
map). For both lobes the degree of polarization was essentially
identical (20 per cent) at the two frequencies; the ratio of degrees
of polarization was 1.0. There is therefore no detection of radio
depolarization in 3C\,223, though we emphasise that the steep spectrum
of the lobes means that there is essentially no detectable 8-GHz lobe
flux density within 45 arcsec of the radio core. We conclude that it
is unlikely that sufficient cold gas is present in the lobes of either
source to allow supersonic lateral expansion.

The presence of sufficient low-energy electrons to provide significant
pressure would require a steepening of the electron energy spectrum
below the observable radio region (e.g. $\gamma < 10^{3}$). Such a
population of electrons cannot be ruled out by our data; however, its
physical origin is unclear. If relativistic protons are providing the
additional pressure, then large proton to electron ratios would be
needed for supersonic expansion. There are arguments suggesting that a
significant proton population is required in parsec-scale jets (e.g.
Celotti \& Fabian 1993), so that they should be present and provide
pressure in the lobes as well. We cannot rule out some contribution
from protons.

In the absence of efficient entrainment, there are few physically
plausible means for the lobes to contain sufficient pressure to be
expanding supersonically in all directions. We cannot rule out
magnetic domination in the lobes or a dominant population of
relativistic protons. However, as discussed above, either of these
scenarios requires the X-ray flux from a power-law model to coincide
with the level expected for minimum energy IC emission in the absence
of protons.

\section{Conclusions}

We have detected X-ray emission from the lobes of two FR-II radio
galaxies, 3C~223 and 3C~284, which in both cases can be attributed to
inverse-Compton emission from cosmic microwave background photons. In
both cases the magnetic field strengths inferred from the levels of
X-ray flux are near to minimum energy values in the absence of
protons. We also detect extended emission surrounding the two sources.
Both sources are in large group-scale atmospheres with temperatures of
$\sim$ 1 keV. Our data are not of sufficient sensitivity to detect
temperature variations in the hot-gas atmosphere. We find that the
lobes are in approximate pressure balance with their external
environments if they contain only relativistic electrons, but that
additional material or magnetic dominance would be required in a model
where the lobes have lateral supersonic expansion.

\section*{Acknowledgments}

We thank the referee for helpful comments. JHC thanks PPARC for a studentship. MJH thanks the Royal Society for a
research fellowship. The National Radio Astronomy Observatory is a
facility of the National Science Foundation operated under cooperative
agreement by Associated Universities, Inc.

\bsp

\label{lastpage}

\end{document}